\documentclass[10pt]{wlscirep}
\usepackage[utf8]{inputenc}
\usepackage[T1]{fontenc}

\title{Wavefront shaping assisted design of spectral splitters and solar concentrators: SpliCons}

\author[1,2,+]{Berk Nezir Gün}
\author[1,2,*]{Emre Yüce}

\affil{Programmable Photonics Group, Department of Physics, Middle East Technical University, 06800, Ankara, Turkey}
\affil[2]{The Center for Solar Energy Research and Applications (GÜNAM), Middle East Technical University, 06800, Ankara, Turkey}

\affil[+]{gun@metu.edu.tr}
\affil[*]{eyuce@metu.edu.tr}

\begin{abstract}

Spectral splitters, as well as solar concentrators, are commonly designed and optimized using numerical methods. Here, we present an experimental method to spectrally split and concentrate broadband light (420 nm - 875 nm) via wavefront shaping. We manage to spatially control white light using a phase-only spatial light modulator. As a result, we are able to split and concentrate three frequency bands, namely red (560 nm - 875 nm), green (425 nm - 620 nm), and blue (420 nm - 535 nm), to two target spots with a total enhancement factor of 715 \%. Despite the significant overlap between the color channels, we obtain spectral splitting ratios as 52 \%, 57 \%, and 66 \% for red, green, and blue channels, respectively. We show that a higher number of adjustable superpixels ensures higher spectral splitting and concentration. We provide the methods to convert an optimized phase pattern into a diffractive optical element that can be fabricated at large scale and low cost. The experimental method that we introduce, for the first time, enables the optimization and design of SpliCons, which is $\sim$ 300 times faster compared to the computational methods.

\end{abstract}
\begin{document}

\flushbottom
\maketitle
\thispagestyle{empty}
\section*{Introduction} Ever since the invention of PhotoVoltaic (PV) cells\cite{doi:10.1063/1.1721711}, the demand for making more efficient solar cells has been the foremost drive for researchers as well as the industry. Scientists constantly seek for effective methods to increase the efficiency of the solar cells by researching more into material science as well as developing methods that involve spatial control of light using solar concentrators\cite{james} and trackers\cite{moore1987solar}. Control over the frequency of incident light opens an additional dimension to benefit from, which provides the means to push the limits beyond a single material can provide. Spectral splitting of incident frequencies provides high PV cell efficiency and cost-effective optical systems\cite{doi:10.1002/pip.988}. Some of the classic spectral beam splitting methods include holographic concentrators\cite{Bloss:82}, thin-film filters\cite{MOJIRI2013654}, prism assisted systems\cite{Stefancich:12} and diffractive optical elements (DOEs)\cite{Dong:96}.

DOEs steer the incoming beam by diffracting it as its name signifies. The impact of DOEs is increasing, specifically in microscopy\cite{DiFabrizio:03} and solar energy research\cite{Mingareev:11}. The DOEs that are employed in solar energy research can spectrally split, concentrate light, or perform these two tasks simultaneously. We think that it is important to distinguish a DOE that can simultaneously perform spectral splitting and concentration and, therefore, we name these special DOEs as SpliCons. DOEs are generally calculated via computational methods such as Gerchberg-Saxton\cite{gerchberg1972practical}, Yang-Gu\cite{Yang:94}, direct binary search\cite{Seldowitz:87}, iterative Fourier transform\cite{10.1117/12.451643}, genetic optimization\cite{Zhou:99} and simulated annealing\cite{Lin:96}. Optimization of DOEs is a rather challenging task if increased level of splitting and concentration efficiencies are targeted. Optimization of a DOE for white light ranging between 400 nm - 1100 nm would require coarsely 700 wavelengths if this range is divided at every 1 nm. A DOE with a lateral pixel number of only 40x40 and with nine steps in pixel height would require 10$^{7}$ parameters to be processed. For this reason, it is computationally expensive to compute a real-life physical system, even with today's improved computational power. The pioneering experimental research by Kim et al.\cite{PhysRevLett.110.123901} has demonstrated spectral splitting of light using a DOE that has 1000 pixels in one dimension. In that study, the heights of the pixels are optimized to spectrally split incident light. The DOEs that are not optimized in 3D cannot fulfill both spectral splitting and concentration of the incident light, which are essential for smart solar cells.

Here, we develop an experimental method to design and optimize SpliCons that can achieve simultaneous concentration and spectral splitting of white light. We use a spatial light modulator (SLM), an electro-optic device that enables programming of holographic phase plates that can emulate a DOE. SLMs that enable dynamic control over the spatial phase of light has significantly contributed to research fields such as optical tweezers\cite{Leach:04}, linear/nonlinear microscopy\cite{doi:10.1002/lpor.200900047}, micro-processing of materials\cite{KUANG20082284}, adaptive optics\cite{10.1117/12.478562}, or pulse shaping\cite{doi:10.1063/1.1150614} and computer-generated holograms\cite{Mok:86}. Here, we employ an SLM for solar energy research and demonstrate spectral splitting and concentration of white light at a record pixel number. The number of pixels we control is $\sim$ 2x10$^{6}$, which is three orders greater than the previous experimental study\cite{PhysRevLett.110.123901}. In our measurement, we use an SLM in order to control the spatial phase of each frequency band. The splitting ratio and the concentration efficiency of a SpliCon increase with the number of controlled pixels. The calculation of a SpliCon that we experimentally optimize here can take up to 89 days using numerical methods, which also lack real-life circumstances\cite{Yolalmaz:20}. The method that we introduce here for solar energy research provides the means to design SpliCons experimentally, enabling $\sim$ 300 times faster optimization compared to the numerical methods.

\section*{Experimental Setup}

\begin{figure}[ht]
\centering
\includegraphics[width=\textwidth]{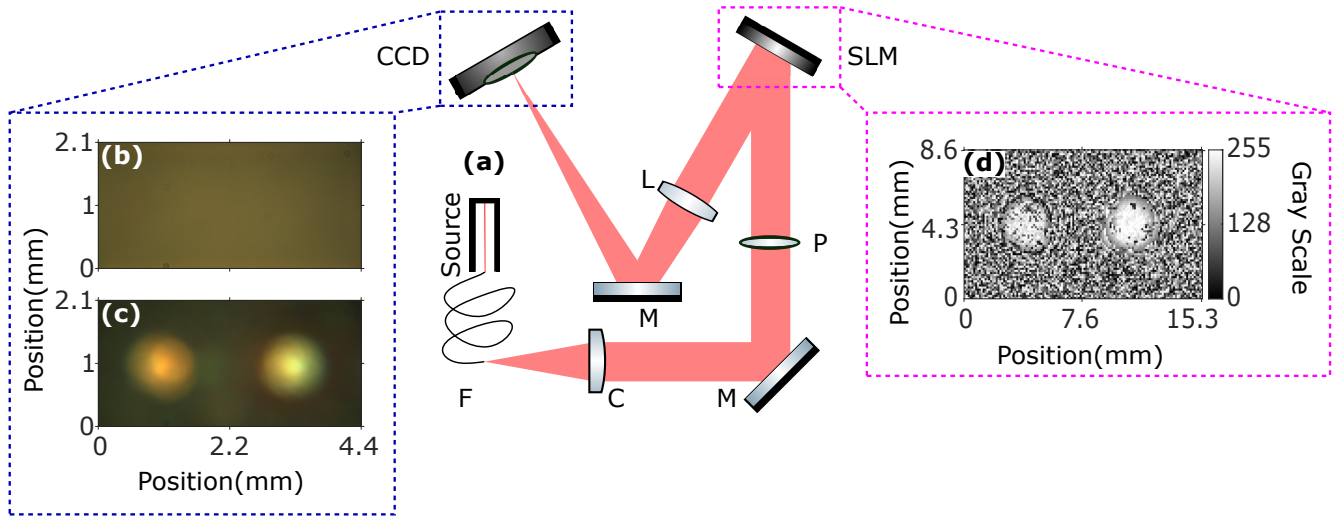}
\caption{(a) The experimental setup that is used to emulate spectral splitters and concentrators: SpliCons using an SLM. Broadband light coming from the source coupled to the fiber (F) passes through condenser lens (C) (f = 160mm) and linear polarizer (P) positioned on the way before SLM. A plano-convex lens (L) (f = 300mm) is placed to transfer the image from the SLM surface onto the camera. (b) Image before optimization of the wavefront. (c) Spectrally split and concentrated light. (d) Phase pattern on the SLM that concentrates and spectrally splits light. The images (b) and (c) are raw images obtained from the CCD camera.} 
\label{fig:Testor3}
\end{figure}

A high degree of temporally incoherent fiber-coupled white light source that emits between 360 nm - 2600 nm was used in our experiment, see Fig. \ref{fig:Testor3}(a). A linear polarizer was positioned after the concentrator lens on the beam path in order to filter the polarization axis that SLM can modulate. We used a phase-only SLM with a nominal resolution of 1920 x 1080 pixels to steer the impinging beam. SLM, whose pixel size is 0.8 $\mu$m, makes an angle of 9.5$^\circ$ with the incoming beam. The SLM that we operated in our experiment can efficiently modulate light between 420 nm - 1064 nm (Holoeye Pluto-2). A converging lens was positioned after the SLM to image the surface of the SLM right before the CCD camera. The multi-channel camera used in our experiment has 1292 x 964 pixels with a pixel size of 3.75$\mu$m. The chip of the color camera (Sony ICX445) has three color channels separated through color filters with the peak quantum efficiencies located at blue: 450 nm, green: 535 nm, and red: 610 nm.

Fig. \ref{fig:Testor3}(b) shows the intensity distribution of light before wavefront shaping. The spectrally split and concentrated light intensity is shown in Fig. \ref{fig:Testor3}(c). The target spots are centered at (1.07 mm, 1.03 mm) and (3.32 mm, 1.03 mm). The intensity increase at the target spots clearly indicates the concentration of the broadband light.  In Fig. \ref{fig:Testor3}(c), it can already be seen that the red portion of the incident light is directed to the left target spot, whereas the green and the blue portion of incident light is directed to the right one. Processed separate color channel data will be presented shortly to illustrate the spectral splitting. Fig. \ref{fig:Testor3}(d) shows the measured phase pattern on the SLM that eventually represents the SpliCon that we optimized.

We measured the optical power of fiber-coupled stabilized tungsten-halogen broadband light by a thermal power sensor (response between 190 nm up to 20000 nm) placed immediately after the condenser, and the average power is 8.49 mW. To detect the manipulated beam, we used a multi-channel camera with an effective chip size of 4.8 mm x 3.6 mm. The active area of phase-only Holoeye Pluto-2 SLM is 15.36 mm x 8.64 mm. Since the SLM surface is larger than the CCD chip, we set the demagnification ratio $\sim$ 3.2 such that the SLM surface is correctly fit on the CCD chip.

\subsection*{Transformation of SpliCons into physical SpliCons}

\begin{figure}[ht]
\centering
\includegraphics[width=.9\textwidth]{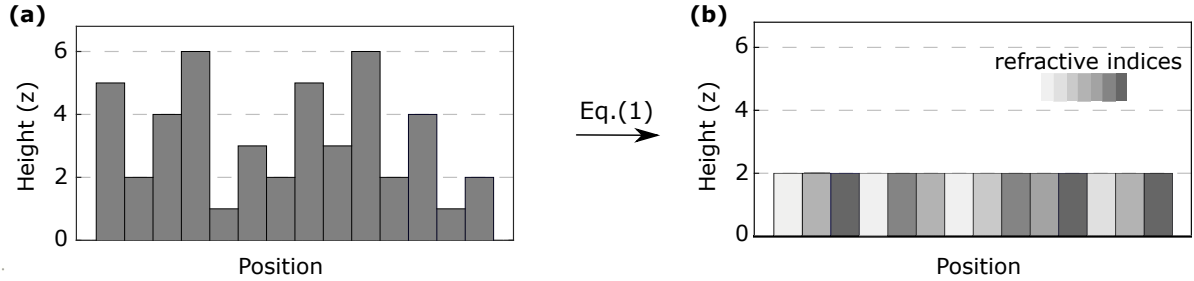}
\caption{(a) An example of physical DOE having seven-step height levels. (b) An example of programmable DOE having seven refractive indices.} 
\label{fig:conversion}
\end{figure}

The use of SLMs for solar panels is, in fact, not realistic due to the cost of SLMs. However, the method that we develop here provide the means to optimize SpliCons in real-life conditions. The measured and optimized phase patterns can easily be converted into a physical SpliCon using the simple conversion of the optical path. The phase that light accumulates through a medium is given by: 

\begin{equation}
\phi_{xy}(\lambda) =  \frac{2\pi}{\lambda} h_{xy}  [n(\lambda)-1] ,
\label{eq:1a}
\end{equation}

\noindent where $\phi_{xy}(\lambda)$ is the phase difference for incoming light on the diffraction plane positioned at (x,y), $h_{xy}$ the thickness of the diffractive material, $\lambda$ the wavelength of the incident light, $n(\lambda)$ the refractive index of the diffractive material. The DOEs are designed and fabricated to control each frequency phase using the variations in $h_{xy}$ in Eq. \eqref{eq:1a}, see Fig. \ref{fig:conversion}(a). On the other hand, the pixels on an SLM have fixed height, and the spatial phase of light is controlled by modifying $n(\lambda)$ in Eq. \eqref{eq:1a}, Fig. \ref{fig:conversion}(b). Each gray index value written on an SLM corresponds to a different refractive index. Using Eq. \eqref{eq:1a} a phase pattern optimized using SLM can be easily transformed into a SpliCon structure that can be fabricated on cost-effective materials at large scales.

\subsection*{Continuous Sequential Optimization Algorithm}

Every pixel on the SLM contributes directly to the image on the camera. However, controlling individual pixels of the SLM does not yield a measurable change on the camera since the change that is introduced by one pixel falls below the noise level\cite{Yilmaz:13}. For this reason, we group the pixels on the SLM and control these linked groups called superpixels. We operate SLM with five particular configurations that increase linearly from 15 x 15 grouped pixels up to 55 x 55 grouped pixels. As a result of the grouping stated, the number of pixels we control on the SLM range from 685 to 9216 in total. Superpixels are fed by using 8-bit gray values from snow white to jet-black. The grayscale on the SLM is divided into 16 steps. Thereby the number of parameters that we can process here reaches up to 9216 x 16 $\cong$ 1.5x10$^{5}$. We have managed to set sub-iteration time within 250 ms. Here, we define a sub-iteration as the scanning of 16 steps in gray indices on a superpixel and measuring the output intensity pattern at the target.

The algorithm's working principle is as follows: after setting the superpixels' size, we encode an entirely white pattern on the SLM, meaning the SLM does not modulate the wavefront since all pixels have the same refractive index. Phase steps for the first group of pixels are encoded on the SLM surface consecutively. Then, we dictate specific target spots on the camera for the concentration of particular frequency bands. The phase step of the specified pixel group that provides the total maximum intensity on the target spots is selected and encoded on the SLM surface. This process is repeated for all superpixels, and a phase pattern is obtained at the end of this major iteration\cite{Vellekoop:07}. We repeat the major iteration three times in order to obtain an optimal phase pattern for the SpliCon. The optimization of the SpliCons takes up to a maximum of 30 hours (for 15 x 15 superpixels with three major iterations) by processing $\sim 4.7x10^{5}$ parameters. The experimental optimization scheme that we introduce here can be decreased to two hours if the integration time on the camera is decreased from 250 ms to 16.6 ms (set by the refresh rate of the SLM). This would require a source and a camera with lower noise levels\cite{Yilmaz:13}.

\section*{Results and Discussion}

\begin{figure}[ht]
\centering
\includegraphics[width=\textwidth]{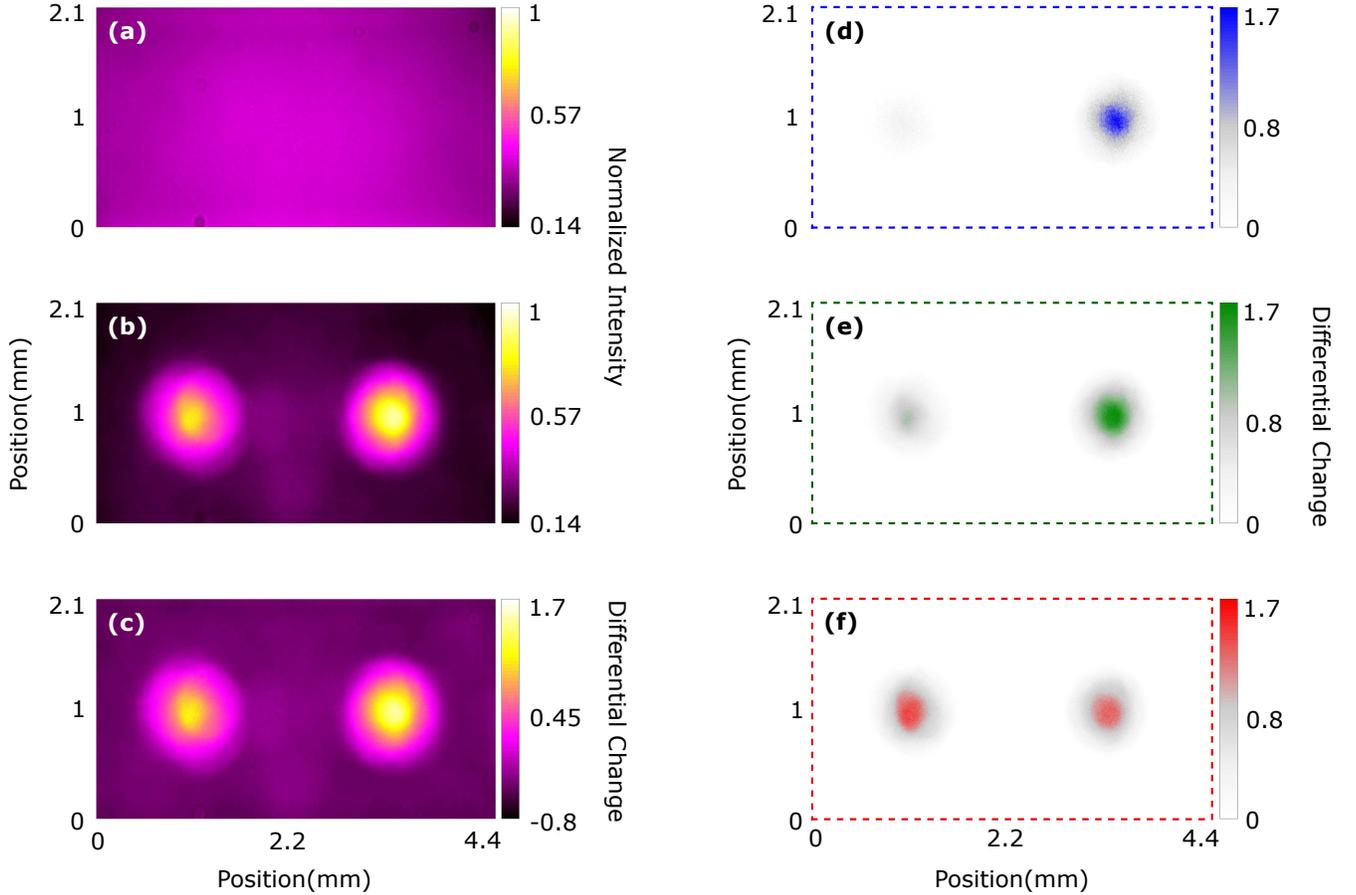}
\caption{The all-in-one single-channel intensity profile (a) before, (b) after wavefront shaping. Differential intensity changes for all-in-one single-channel, blue, green, and red channels are shown separately in panels (c), (d), (e), and (f), respectively.}
\label{fig:Testor4}
\end{figure}

In Fig. \ref{fig:Testor4} we replot the raw images that are shown in Fig. \ref{fig:Testor3}(b) and (c) to quantitatively express the change in intensity for different frequency bands. Fig. \ref{fig:Testor4}(a) and (b) show the intensity profile on camera before and after wavefront shaping, respectively. To illustrate the intensity change, we convert 24-bit multi-channel data provided from CCD camera to 8-bit single-channel data and normalize to the maximum intensity seen in Fig. \ref{fig:Testor4}(b). In Fig. \ref{fig:Testor4}(c) we provide the differential change in intensity that indicates a quantitative measure of the intensity increase due to the optimized SpliCon. The differential change in intensity $\Delta I^{k}(\lambda$) is defined by: 

\begin{equation}{
 \text{$\Delta I^{k}(\lambda$)} = \frac{I^{k}_{f}(\lambda)- I^{k}_{i}(\lambda)}{I^{k}_{i}(\lambda)}, 
  }
\end{equation}

\noindent where k is the pixel number and range between k = \{1,2,..,n\}. I$^{k}(\lambda$) is the scalar-valued function showing spectral intensity values. The subscript i and f stand for the initial and final intensities, respectively.

In Fig. \ref{fig:Testor4} panels (d-f), we provide the differential intensity change for the three particular frequency bands that are between (420 nm - 535 nm), (425 nm - 620 nm), and (560 nm - 875 nm). In our optimization algorithm, we have targeted the beam of light between 420 nm - 620 nm (blue and green bands) to the right spot and the light between 560 nm - 875 nm (red band) to the left spot at the diffraction plane. It can be seen in Fig. \ref{fig:Testor4}(d-f) that the intensities at the targets increase individually for each band, indicating the spectral splitting and concentration of the white light. However, spectral splitting ratios of the blue and the green bands seen in Fig. \ref{fig:Testor4}(d) and (e) have higher values than that of the red band (Fig. \ref{fig:Testor4}(f)). Although we target the red light to the left target position, we observe an increase of red light on the right target position as well. This effect is a natural consequence of CCD's sensing mechanism and stems from the frequency overlap between the bands. At 5 \% quantum efficiency, the red and the green bands of the CCD chip have 29.3 \% overlap, whereas the green and the blue bands have approximately 66.7 \% overlap\cite{guppy}. Nevertheless, we still observe a splitting ratio of S > 0.5 for longer wavelengths.

\begin{figure}[ht]
\centering
\includegraphics[width=\textwidth]{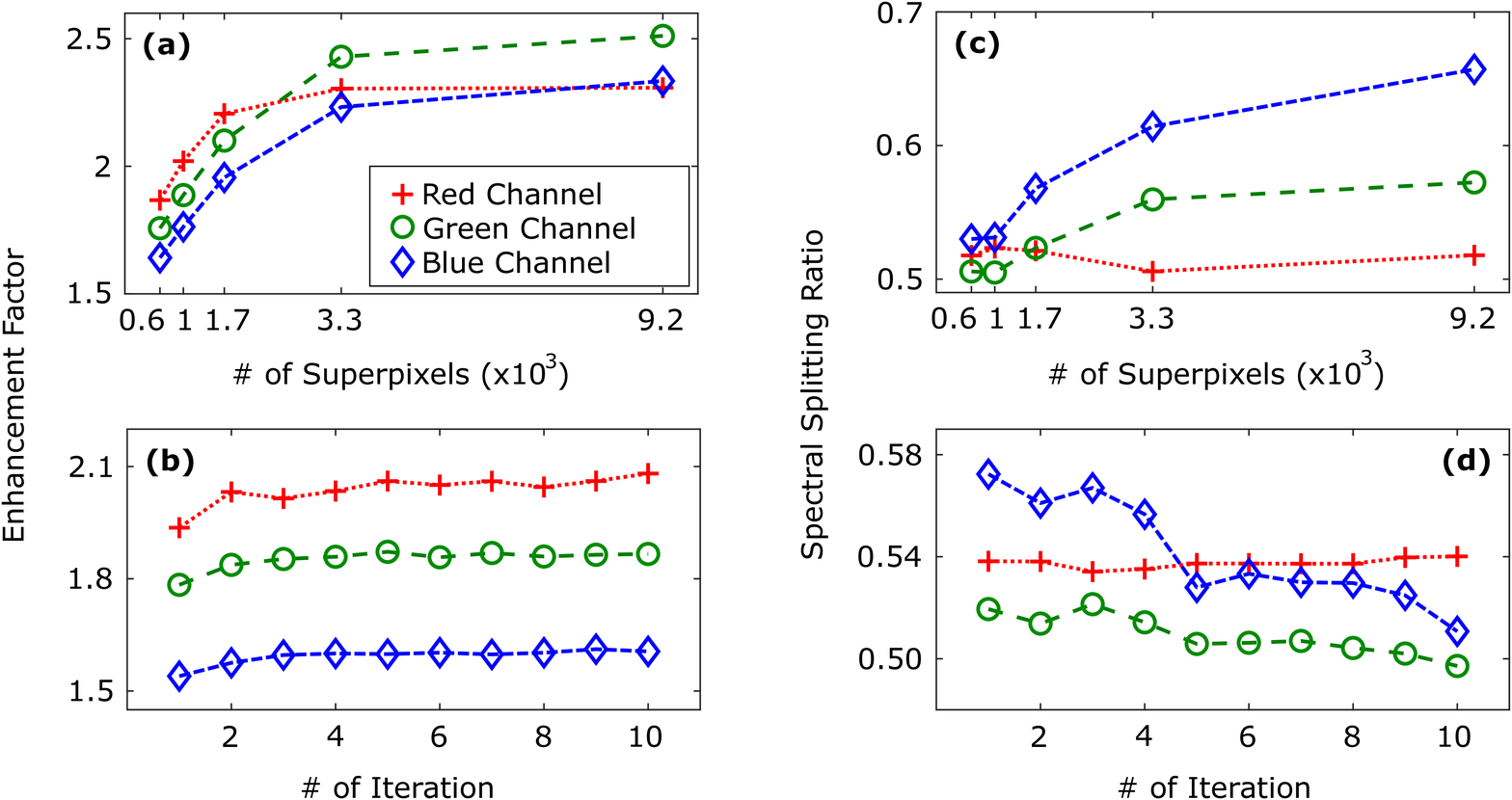}
\caption{The enhancement factor of intensity for each channel as a function of (a) number of superpixels and (b) number of major iterations. The spectral splitting ratio for each channel as a function of (c) number of superpixels (d) number of major iterations. The error bars are smaller than the symbol size in all panels.} 
\label{fig:Testor1}
\end{figure}

\begin{equation}{
\text{$\eta(\lambda)$} = \frac{\int_{T} I_{f}(A_{m},\lambda)\,dA_{m}}{\int_{T} I_{i}(A_{m},\lambda)\,dA_{m}},
} \label{enh}
\end{equation}

\begin{equation}{
\text{S($\lambda$)} = \frac{\int_{T} I_{f}(A_{m},\lambda)\,dA_{m}}{\int_{T} I_{f}(A_{m},\lambda)\,dA_{m}+\int_{T} I_{f}(A_{n},\lambda)\,dA_{n}}, 
} \label{spl}
\end{equation}

We quantify the enhancement factor $\eta(\lambda)$ of the intensity for each channel using Eq. \eqref{enh} in which $A_{m}$ is the area of the m$^{th}$ target spot and m = \{1,2\}. In Eq. \eqref{enh}, $\int_{T} I_{i}(A_{m},\lambda)\,dA_{m}$ and $\int_{T} I_{f}(A_{m},\lambda)\,dA_{m}$ are the spectral intensities on a target spot before and after wavefront shaping, respectively. The enhancement factor is the spectral intensity after the optimization on the corresponding target spot over the incident spectral intensity on the same target spot. We also define the spectral splitting ratio S($\lambda$) in Eq. \eqref{spl}, where m,n = \{1,2\} and m $\neq$ n. It is defined as spectral intensity on the corresponding target spot over total spectral intensity. The letters m and n are assigned to each target spot.

In Fig. \ref{fig:Testor1} we provide data on the effect of the number of superpixels as well as the number of iterations. Through three iterations with 16 distinct gray values, we have obtained the mean of three data sets, which are presented in the panels (a) and (c) in Fig \ref{fig:Testor1}. The enhancement factor in the red channel is the highest for the lower number of superpixels (Fig. \ref{fig:Testor1}(a)). The reason behind that is that the CCD chip has a higher spectral response for the red channel. Less number of adjustable superpixels simply means less control on the light. As the number of superpixels increases, we gain more control over the enhancement factor. The spectral responses provided by the manufacturer\cite{guppy} of the camera are 19.5 x 10$^{-2}$ A/W, 18.0 x 10$^{-2}$A/W and 13.5 x 10$^{-2}$ A/W with the peak wavelengths at 610 nm, 535 nm, 450 nm for the red, green and blue channels, respectively. The enhancement factors of the blue and the green channel intensities surpass that of the red channel intensity for the higher number of adjustable superpixels, given the considerable overlap between the green and the blue channels. Utilizing more superpixels favors more intensity increase at the green-blue target position as a consequence of the algorithm seeking for total maximum intensity. In Fig.\ref{fig:Testor1}(b), we investigate the change in the enhancement factor with respect to the iteration number. We clearly observe that each channel intensity has an increasing trend as the number of iteration increases. The intensity values are saturated at $\sim$ fifth iteration. We have collected the data (Fig. \ref{fig:Testor1}(b) and (d)) by encoding 16 different phase steps to 1024 different superpixels.

The overlaps between the channels have an increasing effect on each other. Unfortunately, these overlaps are disadvantageous for the spectral splitting ratios (Fig. \ref{fig:Testor1}(c) and (d)). By concentrating the green and the blue channel intensities at the same target position rather than defining weights for each channel, we compensate for the higher spectral response for the red channel for the lower number of iterations. Hence, at the green-blue target position, the red channel intensity increases with the increasing number of adjustable superpixels more than the increase in the green and the blue channel intensities at the red target position (Fig. \ref{fig:Testor1}(c)). The decreasing trend of the splitting ratio for the red channel at 3318 superpixels is likely since the major iteration number is not sufficient, and our algorithm favors maximum intensity. If we change the algorithm to get the maximum splitting ratio, we can perhaps observe an increasing splitting ratio with the increasing number of superpixels. Both enhancement factor and spectral splitting ratio values rise with the number of superpixels used for the SpliCon that is programmed on the SLM\cite{vellekoop2008controlling}. In Fig. \ref{fig:Testor1}(d), we use the data for monitoring the spectral splitting ratio as a function of iteration for 1024 adjustable superpixels. After the fourth iteration, the spectral splitting ratio of the red channel becomes dominant due to having the highest spectral response that results in a simultaneous decrease of the splitting ratios of the blue and the green channels. Despite the variations in the spectral splitting ratio, the intensity at each channel increases with the number of superpixels as expected. Defining weights for color channels in the algorithm by considering spectral responses may alleviate the sharp decrease in the blue and the green channel splitting ratios seen at the fifth iteration. However, we predict that even in that case, a smaller overlap between each channel is required, which is not provided by the CCD cameras.

\section*{Conclusion}

By wavefront shaping, we achieved spectral splitting and concentration of broadband light. As an alternative to the conventional computational methods, we employed an experimental approach using an SLM in the design of a SpliCon. By optimizing the gray level values, we optimize refractive indices and achieve simultaneous spectral splitting and concentration of broadband light. The experimentally optimized SpliCon can be transformed into a physical structure on a cost-effective material and could be used in front of the multi-junction solar cells to provide increased efficiency\cite{Huang:13}. The spatial coherence of our source differs from the sunlight to some extend. However, the method that we develop here can be applied to sunlight given the spectral degree of coherence agreement with broadband sources\cite{Divitt:15,Deng,Friberg:95,Mashaal:11,Mashaal:12}. We utilized a particular polarization axis for modulation. Our approach can also be generalized to modulate both polarizations\cite{liu}, which will further increase the enhancement factor and the spectral splitting ratio. In fact, a further increase in the enhancement factor and the spectral splitting ratio can be obtained using an increased number of pixels on the SLM. However, increasing pixel density beyond the large scale fabrication precision limits would not be realistic. For this reason, we keep the pixel size and density achievable with direct laser writing that can provide large scale production of SpliCons\cite{tokel}. We think that the increased speed in optimization time ($\sim$300x), as well as the ability to design SpliCons in real-life conditions, will result in transformative effects in solar cells as they will pave the way for faster optimization of tandem cells with different geometries and material combinations.

\section*{Acknowledgements}

This study is financially supported by The Scientific and Technological Research Council of Turkey (TÜBİTAK), grant no 118F075. We would like to thank Alpan Bek and Allard P. Mosk for providing us the essential equipment. We also would like to thank Raşit Turan and Ahmet Oral for their support during the establishment of our laboratory.

\bibliography{main}

\end{document}